\documentstyle[12pt]{article}
\begin{document}
\baselineskip 15pt

\title{Do dynamical reduction models imply that arithmetic does not apply to 
ordinary macroscopic objects?\footnote{To appear in the March Issue of {\it
Brit. Jou. Phil. Sci.}}}
\author{GianCarlo Ghirardi\footnote{e-mail: ghirardi@ts.infn.it}
\\ {\small Department of Theoretical Physics of the University of Trieste, 
and}\\ {\small the Abdus Salam International Centre for Theoretical Physics, 
Trieste, Italy.} \\ and \\
\\ Angelo Bassi\footnote{e-mail: bassi@ts.infn.it} \\
{\small Department of Theoretical Physics of the University of 
Trieste, Italy.}}
\date{}
\maketitle

\begin{abstract}
We analyze a recent paper in which an alleged devastating criticism 
to the so called GRW proposal to account for the objectification of the 
properties of macroscopic systems has been presented and we show that the 
author has not taken into account the precise implications of the GRW theory. 
This fact makes his conclusions basically wrong. We also perform a survey of 
measurement theory aimed to better focus the physical and the conceptual 
aspects of the so-called macro-objectification problem.
\end{abstract}

\section{Introduction}

Recently, a paper \cite{ghi1}, hereafter referred to as PJL, has been published 
in which an alleged crucial criticism of the so called GRW theory \cite{ghi2} 
has been presented. The basic claim of the paper, which echoes some 
previous analogous criticisms \cite{ghi3} but puts forward new arguments and 
pretends to derive much more drastic conclusions, is that the occurrence
of the so-called ``tails'' of the wavefunctions describing ordinary 
macroscopic objects within the GRW theory, implies that arithmetic 
does not apply to such objects. As a consequence it is argued that 
the GRW approach does not represent an acceptable way out from the 
difficulties of Standard Quantum Mechanics (SQM).

As we will prove in this paper, PJL fails in taking into account the 
precise implications of the GRW theory. This makes all his analysis 
not only physically irrelevant, but conceptually incorrect. After 
having made clear this point in Sections 3 and 4, we will devote the 
following Sections to revisiting some aspects of the measurement 
problem. This seems appropriate because the analysis of the author, 
among other things, involves a misleading mixing up of the real 
``measurement problem of SQM'' with the much more general (but 
universally considered as not puzzling) problem of the description, 
within a genuine Hilbert space formalism, of the continuous degrees of 
freedom of macroscopic systems. Finally, we will point out that some 
well known and mathematically rigorous results show that the author's 
requests cannot be satisfied by almost all measurement processes. 

\section{The argument of PJL.}

The initial remark of ref. \cite{ghi1} is that according to standard quantum 
mechanical dogma mutually exclusive states of affairs must be 
associated to orthogonal vectors, while, within dynamical reduction 
schemes like the GRW theory, one is unavoidably led to release such a 
request for the states of the apparatus corresponding to different 
outcomes. We plainly recognize that the above statement is correct. In 
fact, we all know that the negation of a statement concerning the 
properties possessed by an individual quantum system must be represented, 
within quantum formalism, by orthocomplementation in the Hilbert 
space. We also agree on the fact that in GRW-like theories the role of 
the tails of the wavefunctions is essential to guarantee the desired 
physical implications of such models \cite{ghi4}, and that the presence of the 
tails forbids mutually alternative measurement outcomes being perfectly 
correlated to orthogonal states.

However, as we will prove, the above remark does not allow to draw the 
conclusions of ref \cite{ghi1} that the GRW point of view leads to 
contradictions 
in dealing with the problem of counting a large number of macroscopic 
marbles in a certain space region (and thus it would clash with the 
principles of arithmetic). In fact, the central (and, in the author's 
opinion, fatal) argument of Section 4 of the paper, turns out to be 
wrong due to the fact that the author has not taken into account the 
precise physical implications of the GRW dynamics. 

Let us recall the argument of ref. \cite{ghi1}. The author considers a 
composite system made up of n macroscopic noninteracting marbles, each 
of which is in a state of the type of those characterizing macroscopic 
states within the GRW theory:

\begin{equation}
|\Psi\rangle_{i}\quad =\quad a|\makebox{in}\rangle_{i} + 
b|\makebox{out}\rangle_{i}, \label{ghir1}
\end{equation}

with $|a|^{2} \gg |b|^{2} \neq 0, |a|^{2} + |b|^{2} = 1$. 
The states $|\makebox{in}\rangle_{i}$ and $|\makebox{out}\rangle_{i}$
correspond to wavefunctions having their support
strictly within an extremely large ``box'' B, and strictly outside it, 
respectively. The box is supposed to be so large that the various indices {\it
i} correspond to locations of the marbles so far from each other that (in the 
author's words) {\it we can keep their interactions as small as we like}.

	Then, according to PJL, the difficulty that the GRW approach meets and 
which should be fatal for it can be easily exhibited:

\begin{enumerate}
\item The GRW theory asserts  that when the i-th particle is in the state
(\ref{ghir1}) ``it counts as being within the box''.

\item Since all {\it n} particles are in a state of this kind, one can 
claim: ``particle 1 is within the box, particle 2 is within the box, ..., 
particle {\it n} is within the box''. 
But, if arithmetic is true, i.e., if the enumeration principle holds, 
this amounts to claim that all particles are in the box.

\item On the other hand, the state of the composite system is:
\begin{equation}
|\Psi\rangle_{all}\quad =\quad [a|\makebox{in}\rangle_{1} + 
b|\makebox{out}\rangle_{1}]\otimes [a|\makebox{in}\rangle_{2} + 
b|\makebox{out}\rangle_{2}]\otimes \ldots\otimes [a|\makebox{in}\rangle_{n} + 
b|\makebox{out}\rangle_{n}].
\label{ghir2}
\end{equation}

\item For such a state quantum mechanics predicts that the probability of 
finding 
all particles within the box in a measurement is given by the square of the 
modulus of the coefficient of the term $|\makebox{in}\rangle_{1}\otimes 
|\makebox{in}\rangle_{2}\otimes\ldots\otimes |\makebox{in}\rangle_{n}$
in the above equation, i.e. by the quantity $|a|^{2n}$. Now, since $|a|^{2}$, 
even though it is extremely close to 1 is actually smaller than 1, for 
sufficiently large {\it n} such a probability becomes appreciably 
smaller than one; actually, it tends to zero for {\it n} tending to infinity.

\item The inconsistency should now be evident: the GRW's claim that when the 
state 
of the {\it i}--th particle is (\ref{ghir1}) the particle is 
within the box and the assumption 
that the enumeration principle of arithmetic holds lead to the assertion ``all 
particles are in the box'', while the outcome of an experiment will (almost) 
surely contradict such a statement. The conclusion follows: accepting GRW 
requires abandoning arithmetic.
\end{enumerate}

\section{Taking into account the physical implications of the GRW dynamics.}

In this Section we will show that above argument is basically incorrect.
This can be easily proved by investigating in greater detail the 
physical implications of the ``crucial'' state (\ref{ghir2}), the one which, in 
the author's opinion, leads to a contradiction between the GRW 
position and the validity of the enumeration principle\footnote{Note that 
here we do not question the assumption of PJL that the number {\it n} of 
marbles be extremely large (a necessary condition in order that his argument 
hold --- see however the comments of the next Section) and we plainly accept 
that
the value of $|a|^{2n}$ turns out to be close to zero.}. The crucial 
point is that the author has not taken into account what the GRW theory
entails for a state like (\ref{ghir2}).
To develope our proof, let us begin by rewriting this state:

\begin{equation}
|\Psi\rangle_{all}\quad =\quad [a|\makebox{in}\rangle_{1} + 
b|\makebox{out}\rangle_{1}]\otimes [a|\makebox{in}\rangle_{2} + 
b|\makebox{out}\rangle_{2}]\otimes \ldots\otimes [a|\makebox{in}\rangle_{n} + 
b|\makebox{out}\rangle_{n}],
\label{ghir3}
\end{equation}

and let us express it as the linear superposition of the states obtained by 
performing explicity the product of the various factors appearing in it. We get
 an extremely long succession of terms, all with extremely small coefficients:

\begin{eqnarray}
& & a^{n}|\makebox{in}\rangle_{1}\otimes |\makebox{in}\rangle_{2}\otimes
\ldots\otimes |\makebox{in}\rangle_{n} +a^{n-1}b|\makebox{out}\rangle_{1}\otimes
|\makebox{in}\rangle_{2}\otimes\ldots\otimes |\makebox{in}\rangle_{n}
\nonumber \\ & &
+ a^{n-1}b|\makebox{in}\rangle_{1}\otimes |\makebox{out}\rangle_{2}\otimes\ldots
\otimes |\makebox{in}\rangle_{n} +\ldots + a^{n-1}b|\makebox{in}\rangle_{1}
\otimes |\makebox{in}\rangle_{2}\otimes\ldots\otimes |\makebox{out}\rangle_{n}
\label{ghir4} \\ & &
+a^{n-2}b^{2}|\makebox{out}\rangle_{1}\otimes |\makebox{in}\rangle_{2}\otimes
\ldots\otimes |\makebox{in}\rangle_{n} +\ldots +b^{n}|\makebox{out}\rangle_{1}
\otimes |\makebox{out}\rangle_{2}\otimes\ldots\otimes |\makebox{out}\rangle_{n}.
\nonumber
\end{eqnarray}

With reference to this expression one has to take into account two fundamental 
facts:

\begin{enumerate}

\item[i)] Any pair of terms in (\ref{ghir4}) corresponds to 
different locations of at least one 
marble: in the first term all marbles are within the box and in different 
positions (so as to guarantee that {\it their interactions
are as small as we like}),
in the second term marble number 1 is outside the box, and so on for the 
remaining states.

\item[ii)] The marbles are macroscopic objects, and, as such, 
they contain a number of 
particles of the order of Avogadro's number. But it is {\bf the most fundamental 
physical characteristic} of the GRW theory that it forbids the persistence of 
superpositions of states of this kind. In particular for the case under 
consideration the precise GRW dynamics will lead {\bf in about one millionth of 
a 
second} to the suppression of the superposition and to the ``spontaneous 
reduction'' of the state (\ref{ghir4}) to one of its terms (with the probability 
attached to it by its specific coefficient). Thus, the difficulty arising, 
according to PJL, in connection with a state like 
(\ref{ghir4}) does not present itself 
for the simple reason that such a state never occurs, its existence being 
forbidden by the GRW theory itself. Accordingly, the statement that the GRW 
approach leads to claim that such a state describes the situation in which 
``precisely {\it n} particles are within the box'' 
is nonsensical for the very simple 
reason that such a state is not a possible state for the considered physical 
system of n marbles.
\end{enumerate}
 
	Summarizing, the situation that PJL envisages to show that the 
dynamical reduction program meets extremely serious difficulties, 
involves a state which, according to the theory he is criticizing, 
cannot occur and, even if it would occur at a given instant, it would 
be transformed immediately into a pefectly reasonable (from the point 
of view of the enumeration principle) state in which {\bf an absolutely 
precise number} of marbles is within the box. Obviously, there is a 
non-zero probability that this number be different from n, but this is 
totally irrelevant: the GRW theory, while claiming that a marble in 
the state (\ref{ghir1}) can be asserted ``to be within the box'', due to its 
basic dynamical features has precise physical implications which PJL 
seems to ignore concerning hypothetical situations in which more than 
one marble is in such a state.\footnote{It seems of some relevance to us 
to call the attention of the reader on the 
analysis of Section 3.4 of ref. \cite{ghi5} in 
which a detailed comparison of Hilbert 
space criteria and average mass density criteria to distinguish macroscopically
different situations is done. Obviously, the mass density criteria do not 
associate strictly exclusive alternatives to orthogonal statevectors. B. van 
Fraassen, in commenting the approach and criteria adopted in ref. \cite{ghi5} 
to characterize macroscopic differences has stated (private correspondence):{\it 
I have talked with my students about your paper and also brought it up in David 
Albert's seminar. We all agreed that your paper addresses the most important 
issue about how to relate quantum mechanics to the macroscopic phenomena in a 
truly fundamental and new way.}  And he also wrote: 
{\it It seems to me that we are 
seeing a consensus being reached, which cuts across several rival approaches, 
on how to handle the problem of aligning macroscopic with microscopic 
descriptions. This is to say that such reports as that a real measurement had 
a certain outcome, that the pointer was at the 7 for instance, are not to be 
equated with statements to the effect that a certain specific observable had 
a specific value (or that a certain vector was an eigenvector of a specific 
observable) but rather with something else which - some vagueness apart - can 
still be described in quantum mechanical terms. ... I will explain below why 
I see this as part of a consensus with discussions about other 
interpretations of quantum mechanics. But there is this difference: that you 
have given, in your discussion of appropriate and inappropriate topologies, 
an important and even (to my mind) very convincing rationale for this solution.}
It is conforting for us that a philosopher deeply involved in the 
foundational issues of quantum mechanics sees in our analysis a {\it very 
convincing rationale} for abandoning the requirement of strict orthogonality 
of the final states of the apparatuses and not a reason to claim that this 
makes the GRW proposal to conflict with basic requirements of arithmetic.} 
When these implications are correctly 
taken into account it emerges clearly that it is meaningless to make 
any statement about the location of the marbles in such states simply 
because they cannot persist for \cite{ghi6} ``more than a split second''. 
Actually, even in the case of only two marbles the theory implies that it is 
nonsensical to consider a state like: 

\begin{equation}
|\Psi(1,2)\rangle\quad =\quad [a|\makebox{in}\rangle_{1} + 
b|\makebox{out}\rangle_{1}]\otimes [a|\makebox{in}\rangle_{2} + 
b|\makebox{out}\rangle_{2}]
\label{ghir5}
\end{equation}

and, to grasp its physical implications, requires to make reference to the non 
puzzling terms of the expression: 

\begin{eqnarray}
|\Psi(1,2)\rangle & = & a^{2}|\makebox{in}\rangle_{1}\otimes
|\makebox{in}\rangle_{2} + ab|\makebox{out}\rangle_{1}\otimes
|\makebox{in}\rangle_{2} \nonumber \\
& + & ab|\makebox{in}\rangle_{1}\otimes |\makebox{out}\rangle_{2} +
b^{2}|\makebox{out}\rangle_{1}\otimes |\makebox{out}\rangle_{2}
\label{ghir6}
\end{eqnarray}

one of which becomes immediately (i.e. within $10^{-6}$--$10^{-7}$ sec.) 
the one describing the actual physical situation.

	The above analysis shows that the criticism of PJL is devoid of any 
sense.\footnote{We stress that, even though we have chosen
to develop our argument with 
reference to the simplified GRW version of the spontaneous localization 
mechanism, precisely the same argument holds for the continuous versions of 
dynamical reduction models, in particular CSL, which take into account the 
identity of the constituents. In such models, as discussed in great detail in 
ref.[7], instead of speaking of the number of particles and of their almost 
precise locations one should make reference to the appropriately averaged 
local mass density at all points of ordinary space.}

\section{A quantitative analysis.}

With reference to the just discussed aspect of the theory which implies
that a state like (\ref{ghir3}) is transformed ``in a split second'' into one of
the terms of its expression (\ref{ghir4}), we consider it illuminating to 
take precisely into account the details of the GRW dynamics to 
evaluate the various possible outcomes of the spontaneous reduction 
process. To this purpose let us recall that in ref. \cite{ghi5} consideration 
has been given to a system which is initially in the less localized 
situation (from the point of view of the present analysis) which can 
occur, i.e. in the state:

\begin{equation}
|\Psi\rangle_{non loc}\quad =\quad\frac{1}{\sqrt{2}}[|\makebox{in}\rangle +
|\makebox{out}\rangle ], \label{ghir7}
\end{equation}

and it has been proved that after a time t of the order of the perception time 
(i.e. $10^{-2}$--$10^{-3}$ sec) it is dynamically 
transformed by the universal GRW evolution equation 
in a state like (\ref{ghir1}). Moreover, in the case of a macroscopic system, a 
rigorous realistic explicit estimate of the quantities 
$|a|^{2}$ and $|b|^{2}$, after the above 
time has elapsed, has been given. The argument of ref.
\cite{ghi5} implies that, in the 
case of a macroscopic body of normal density and with a volume of one cubic 
centimeter (i.e. for a system like the marbles of PJL) one has, after the 
considered time:

\begin{equation}
|b|^{2} \cong e^{-2(10^{15})} \cong 10^{-(10^{15})}.
\label{ghir8}
\end{equation}

Let us then take once more into consideration the state (\ref{ghir3}), let us 
define the operator  which counts the particles in the box and let us 
evaluate explicitly the probability  $P(\hat{N}_{in} = n|\Psi_{all})$
that, as a consequence of the 
reduction process which it suffers almost immediately, it is 
transformed into the first term of Eq. (\ref{ghir4}), i.e. in the state 
corresponding to the property ``all n particles are within the box''. 
Such a probability is given by:

\begin{equation}
P(\hat{N}_{in} = n|\Psi_{all}) = |a|^{2n} =
(1 - |b|^{2})^{n}. \label{ghir9}
\end{equation}

This equation implies that the probability  $P(\hat{N}_{in} \neq n|\Psi_{all})$
that the reduction process leads to any other state appearing in
(\ref{ghir4}) is:

\begin{equation}
P(\hat{N}_{in} \neq n|\Psi_{all}) = 1 -
(1 - |b|^{2})^{n}. \label{ghir10}
\end{equation}

We choose now a positive number $\tau$ appreciably smaller than 1 and we 
determine 
for which value of n the probability  $P(\hat{N}_{in} \neq n|\Psi_{all})$
takes a value larger than $\tau$. We have:

\begin{equation}
1 - (1 - |b|^{2})^{n} > \tau, \label{ghir11}
\end{equation}

i.e.:

\begin{equation}
(1 - |b|^{2})^{n} < 1 - \tau \label{ghir12}
\end{equation}

This equation amounts to:

\begin{equation}
n > \frac{\ln [1 - \tau]}{\ln [1 - |b|^{2}]}, \label{ghir13}
\end{equation}

which, by taking into account that $\tau$ and  $|b|^{2}$
are extremely small, gives

\begin{equation}
n > \frac{\tau}{|b|^{2}},\,\,\,\makebox{i.e.}\,\,\, \tau < n|b|^{2}.
\label{ghir14}
\end{equation}

Eq. (\ref{ghir14}) has some extremely interesting implications: \\ \\
--- Let us suppose that the number of marbles we want to consider is so large 
that their mass equals the mass of the whole universe, which, as is well known,
is of the order of $10^{53}$ gr.
Assuming that a marble has a mass of the order of 1 gr,  $n$
turns out to be of the order of $10^{53}$. 
Taking into account the value (\ref{ghir8}) and Eq. (\ref{ghir14}) we get:

\begin{equation}
\tau \leq 10^{53}\cdot 10^{-(10^{15})} \cong 10^{-(10^{15})}. \label{ghir15}
\end{equation}

The conclusion is then obvious: {\bf even if one were able}  (but the 
theory makes this practically impossible) to prepare at a given instant
a state like (\ref{ghir1}) for which $n$ is so large that one must use the mass 
of the whole universe to build the marbles which appear in it, it 
would immediately collapse onto one of its terms and the probability 
that this term corresponds to a number of particles in the box 
{\bf different in any way whatsoever} from $n$ is smaller than 
$10^{-(10^{15})}$.

We do not want to be misunderstood: we know very well that one cannot 
call into play the smallness of the probability of a certain 
occurrence to avoid a mathematical contradiction. But the purpose of 
the above calculation is not that of escaping from a contradiction. As 
it should be perfectly clear to everybody who has followed us, no 
logical problem arises from the possibility that the reduction leads 
to any state of Eq.(\ref{ghir4}) different from its first term. We have 
presented the explicit evaluation of the probabilities of the various 
reductions simply to make clear to the reader the actual physical 
orders of magnitude which govern which one of the potential states of 
(\ref{ghir4}) is actualized by the GRW mechanism and that such a state is 
(almost) certainly the one corresponding to n particles being within 
the box.

\section{The orthogonality requirement for the final 
states in a measurement process.}

We consider it appropriate to devote the second part of this paper to 
critically analyze the requirement that the final apparatus states, 
after the measurement process is over, be strictly orthogonal. This 
will allow us to focus on the difference between the ``quantum 
measurement problem'' and the problem of describing the continuous 
degrees of freedom of a macroscopic object.

\subsection{General considerations.}

We begin by pointing out that the problem which is raised by the 
appearance of the tails in theories of the GRW--type has very little to 
do with the so-called ``measurement problem of SQM'' but instead it is 
strictly related to another relevant problem, i.e. the one of 
describing, within a genuine Hilbert space formalism (i.e. without 
resorting to some kind of hidden variables), a physical system having 
a definite location. To see this we can start by recalling that von 
Neumann, in his fundamental attempt to account in a mathematically 
precise way for the measurement problem, has considered an idealized 
situation in which a microsystem $S$, in one, let us say 
$|\omega_{j}\rangle$, of the proper eigenstates of a self--adjoint 
operator $\hat{\Omega}$ corresponding to the physical 
observable $O$  we are interested in, interacts with a macroscopic 
apparatus $A$ in a ``ready to measure'' state  
$|A_{0}\rangle$ and has hypothesized a 
precise dynamics governing the $S$--$A$ interaction leading to the evolution:
	
\begin{equation}
|\omega_{j}\rangle\otimes |A_{0}\rangle\quad\Longrightarrow\quad
|\omega_{j}\rangle\otimes |A_{j}\rangle. \label{ghir16}
\end{equation}

The states $|A_{j}\rangle$, i.e. those accounting 
for the amplification from the 
``unaccessible'' microworld to the world of our definite perceptions, are 
assumed to satisfy:

\begin{equation}
\langle A_{k}|A_{j}\rangle = \delta_{kj}. \label{ghir17}
\end{equation}

Such relations are obviously necessary if one wants to infer {\it with absolute 
precision}, from a subsequent measurement on the apparatus, the property 
referring to the observable $O$ which is possessed by the microsystem. In fact, 
would the scalar product $\langle A_{k}|A_{j}\rangle$
be different from zero for $k \neq j$ , one would not be allowed
to deduce from the outcome $a_{k}$ of the test he performs on the 
apparatus that the system has been found in the state $|\omega_{k}\rangle$ 
in the measurement: actually, it could have been found in the state 
$|\omega_{j}\rangle$. 

Let us now suppose that all previous assumptions about the measurement 
process are satisfied. Then one can formulate in a simple way the so 
called measurement problem: since the system S can be (in general) 
easily prepared in a linear superposition of the states $|\omega_{k}\rangle$:

\begin{equation}
|\Psi^{(S)}\rangle = \sum_{k} c_{k}|\omega_{k}\rangle, \qquad
\sum_{k} |c_{k}|^{2} = 1. \label{ghir18}
\end{equation}

Eqs. (\ref{ghir16}) and the natural assumption that the linear evolution 
of SQM governs all physical processes, imply:

\begin{equation}
|\Psi^{(S)}\rangle\otimes |A_{0}\rangle\quad\Longrightarrow\quad 
\sum_{k} c_{k}|\omega_{k}\rangle\otimes |A_{k}\rangle. \label{ghir19}
\end{equation}

But then: what meaning can be attached to a final statevector like the one at 
the r.h.s. of Eq.(\ref{ghir19}) in which states corresponding to 
different macroscopic situations of the  apparatus appear?

	In our opinion it is of remarkable relevance to stress that both in 
the specific theoretical models of measurement processes worked out by 
von Neumann himself \cite{ghi8} as well as in practically all actual 
experimental situations, one is led to consider the states $|A_{j}\rangle$ as 
referring (according to the index which characterizes them) to 
different possible spatial locations of a macroscopic object (the 
pointer of the apparatus). This remark actually makes more striking 
(as repeatedly stressed by Schr\"{o}dinger \cite{ghi9} and Einstein 
\cite{ghi10}, 
among many others) the embarrassment with a state like (\ref{ghir19}): how can a 
macroscopic system be in a superposition of states corresponding to 
its being located in macroscopically different spatial regions? The 
situation is so puzzling that SQM has tried to circumvent it by 
adopting the point of view that J.S. Bell has appropriately 
characterized \cite{ghi5} as the assumption that 
{\it Schr\"{o}dinger's equation is not 
always right}: in the considered process, since a macroscopic system 
enters into play, the linear and deterministic Q--evolution breaks 
down, the nonlinear and stochastic process of wave packet reduction 
(WPR) occurs, and the final state reduces to one of the terms of the 
sum at the r.h.s. of (\ref{ghir19}), with probability 
$|c_{k}|^{2}$. The GRW theory gets 
this result by assuming, in place of an ill defined postulate (WPR) 
which contradicts the basic principles of the theory, that the 
dynamics of SQM has to be modified in such a way that one can derive 
the desired behaviour from a unified dynamical law governing all 
physical processes, from the microscopic to the macroscopic scale. 
Such a procedure has been proved to be viable but it requires two 
precise moves: 

-- One must take extremely seriously the idea that the information we can get 
about the states of a microsystem must be correlated to the position 
(or better, in the more sophisticated versions of dynamical reduction models 
\cite{ghi5}, to the appropriately averaged mass distribution) of some 
macroscopic 
system.

-- One must be satisfied with the fact that the final statevectors 
$|A_{j}\rangle$ which are 
brought in by the modified dynamics are almost, but not exactly, orthogonal. 
To prepare the basis for the detailed analysis of the next subsection, let us 
consider the simplified version of Eq.(\ref{ghir19}) in which only two 
microscopic 
states appear, let us disregard the internal variables of the apparatus and 
let us assume that, for the purposes we are interested in, the apparatus 
itself is adequately described by a one dimensional configuration variable $X$ 
to 
be identified with the coordinate of its centre of mass. 
Then, denoting as $\Psi_{k}(X) = \langle X|A_{k}\rangle$, $k = 1,2$, 
the final wavefunctions which the GRW theory attaches to the permitted 
and alternative final states of the pointer, one immediately realizes that
$\Psi_{1}(X)$ and $\Psi_{2}(X)$
have a nonvanishing (even though extremely small) overlap and, in general,
they are not exactly orthogonal.

\subsection{The von Neumann dynamical model of measurement and, 
more generally the wavefunction of macrosystems.}

This Section will be devoted to make clear why one has to keep quite 
distinct the measurement problem from the one of accounting for a 
macro-object having a precise location. Actually, von Neumann himself 
\cite{ghi8}, (as well as many other authors) has described the measurement 
process by making precise reference to a model in which the apparatus 
$A$ is, once more, described by its centre-of-mass position variable $\hat{X}$. 
The initial apparatus state, in the coordinate representation, is a 
Gaussian of width $\delta$ centred around the position $X = 0$:

\begin{equation}
\langle X|\Psi_{0}\rangle\equiv\Psi_{0}(X) = \left(\frac{1}{\delta
\sqrt{2\pi}}\right)^{\frac{1}{2}} e^{- \frac{X^{2}}{4\delta^{2}}}.
\label{ghir20}
\end{equation}

Such a system interacts with the microscopic system $S$ for which we are 
interested in determining the value of the observable $O$. One disregards the 
free motion both of the microsystem and of the apparatus and assumes the 
interaction between $S$ and $A$ to be governed by the following hamiltonian:

\begin{equation}
H = -\gamma\hat{\Omega}\hat{P}, \label{ghir21}
\end{equation}

where $\gamma$ is a coupling constant with the appropriate dimensions,
$\hat{\Omega}$ is the operator corresponding to $O$, 
and $\hat{P}$ is the momentum canonically conjugated to the position 
variable $\hat{X}$. Let us assume that the interaction is switched on for 
a time interval $T$. The evolution, when the system $S$ is in one of the two 
eigenstates of $\hat{\Omega}$, is described by the relations:

\begin{eqnarray}
& & |\omega_{1}\rangle\otimes |\Psi_{0}\rangle\quad\Longrightarrow\quad
|\omega_{1}\rangle\otimes |\Psi_{1}\rangle \nonumber \\
& & |\omega_{2}\rangle\otimes |\Psi_{0}\rangle\quad\Longrightarrow\quad
|\omega_{2}\rangle\otimes |\Psi_{2}\rangle, \label{ghir22}
\end{eqnarray}

where:

\begin{eqnarray}
\Psi_{1}(X,T)\quad =\quad \Psi_{0}(X - \gamma\omega_{1} T) \nonumber \\
\Psi_{2}(X,T)\quad =\quad \Psi_{0}(X - \gamma\omega_{2} T). \label{ghir23}
\end{eqnarray}

Then, if one assumes that $\gamma (\omega_{2} - \omega_{1})T \gg \delta$
one claims that ``the fact that the final pointer position is around $\gamma
\omega_{1}T$ $(\gamma\omega_{2}T)$ tells us that the initial 
state of the system was $|\omega_{1}\rangle$ $(|\omega_{2}\rangle)$''. 
In other words, one has been able to devise an apparatus amplifying the 
initial microproperty of the system, correlating it to one of two 
macroscopically different locations of a macroscopic object.

	In the considered formal description the quantum measurement problem 
consists in the fact that triggering the apparatus in the same initial 
state with the microstate:

\begin{equation}
|s\rangle\quad =\quad\frac{1}{\sqrt{2}} [|\omega_{1}\rangle + 
|\omega_{2}\rangle ] \label{ghir24}
\end{equation}

yields a final entangled state involving the superposition of 
the states $\Psi_{1}(X,T)$ and $\Psi_{2}(X,T)$ of Eq.(\ref{ghir23}). 
We stress that neither von Neumann, nor all other scientists 
discussing this scheme, have ever considered as puzzling the appearance 
of the states $\Psi_{1}(X,T)$ and  $\Psi_{2}(X,T)$
at the r.h.s. of Eq.(\ref{ghir22}) but only the occurrence of their 
superposition 
when the initial state of the system is the one given by (\ref{ghir24}).

	At this point it is appropriate to discuss in greater detail the basic 
features of a state like those of equation (\ref{ghir23}). For simplicity we 
will assume it to be centred around the origin, so that we will 
actually deal with the state (\ref{ghir20}). It is obvious that such a state, 
which in the case of an appropriately (on the relevant scale) small $\delta$
is interpreted by everybody as describing a pointer centred at $X = 0$, has 
the following properties:

--- If we consider an arbitrarily large space interval $(-D, D)$
and we call $|\makebox{in}\rangle$ and $|\makebox{out}\rangle$ two 
normalized states such that:

\begin{equation}
\langle X|\makebox{in}\rangle\quad =\quad\left\{ \begin{array}{cl}
\frac{\langle X|\Psi_{0}\rangle}{\sqrt{N_{in}}}, & \makebox{when 
	$X \in (-D, D)$} \\
0, & \makebox{otherwise} \end{array} \right. \label{ghir25}
\end{equation}

\begin{equation}
\langle X|\makebox{out}\rangle\quad =\quad\left\{ \begin{array}{cl}
0, & \makebox{when $X \in (-D, D)$} \\
\frac{\langle X|\Psi_{0}\rangle}{\sqrt{N_{out}}}, & 
\makebox{otherwise}, \end{array} \right. \label{ghir26}
\end{equation}

one has: 

\begin{equation}
|\Psi_{0}\rangle\quad =\quad\sqrt{N_{in}} |\makebox{in}\rangle +
\sqrt{N_{out}} |\makebox{out}\rangle, \label{ghir27}
\end{equation}

with:

\begin{equation}
N_{in} \gg N_{out}. \label{ghir28}
\end{equation}

	Concluding: the state which everybody would associate to the assertion 
``the pointer points at 0'' has exactly the same features of the state 
(\ref{ghir1}) of this paper i.e. of the state (4) of ref.\cite{ghi1}, the 
typical 
final state of the GRW theory, the one which, according to PJL, should 
have unacceptable implications. In particular, no matter how large is 
taken the ``box'' $(-D ,D)$, and how well localized is  
$\Psi_{0}(X)$ (i.e. how small is $\delta$), the 
state attributes a non exactly zero probability to the pointer being 
found outside the ``box''.

	It is to some extent amusing to remark that the criticism of PJL, which
is directed against the GRW's proposal, when considered in the 
perspective we have just outlined, becomes an argument against the 
orthodox interpretation of quantum mechanics. In fact, as we have 
pointed out, all supporters of this point of view including von 
Neumann himself plainly accept that the state (\ref{ghir1}) corresponds to a 
precise location of the macroscopic pointer. On the other hand, 
within SQM the state (\ref{ghir2}) is certainly allowed and moreover it is 
stable and can persist for extremely long times (since the 
wavefunctions of macroscopic bodies spread very little), so that it 
is precisely against such a formalism that one could raise a 
criticism of the type of the one of ref. \cite{ghi1}.

	It should be absolutely obvious that the situation presently discussed 
is not due to having chosen an inappropriate wavefunction to account 
for the situation ``the pointer points at 0'': it is simply an unavoidable
consequence of the fact that, within SQM, no wavefunction can have a 
compact support in configuration space except that for a precise time 
instant t. In fact, for an isolated system, the wave function of the 
centre of mass obeys the Schr\"{o}dinger equation for a free particle, 
implying that, even if at $t = 0$ its support is entirely contained in 
$(-D, D)$, at any subsequent time it extends over the whole real axis. 

	The relevance of these remarks for the argument of PJL should be 
obvious: if he wants to stick strictly to the requirement that the 
final apparatus states are orthogonal, then he must assume that the 
only physical systems which are acceptable as measuring instruments 
are those for which ``the pointer variable'' has a discrete spectrum and 
whose initial state is driven in the closed linear manifolds\footnote{Obviously, 
this is not the whole story: one has still to devise a mechanism 
leading to the objectification of only one of such properties when the initial 
state is a superposition of states belonging to different values for the 
measured observable.}
associated to different discrete eigenvalues according to the 
different microstates of S triggering the process. But such a request 
is certainly rather peculiar and does not correspond to any real 
measurement situation in our laboratories, and, as far as we know, to 
the ways in which our definite perceptions about macroscopic objects 
emerge.

\subsection{A general theorem about measurements.}

	As previously stated, there are other, more formal and rigorous ways 
of proving that the requests of PJL cannot be satisfied, which are 
related to the so called Wigner-Araki-Yanase theorem
\cite{ghi11, ghi12, ghi13, ghi14} putting 
precise limitations to the very possibility of ideal measurements for 
almost all observables. Here we will present them in the very sketchy 
way discussed in the introduction of ref.\cite{ghi13}. Let $S$ be, as usual, the 
measured system and $A$ the measuring apparatus. Let 

\begin{equation}
\hat{M}|m\rangle\quad =\quad m|m\rangle \label{ghir29}
\end{equation}

be the eigenvalue equation for the observable of $S$ we are interested in 
measuring. Let $\hat{\Gamma} = \hat{\Gamma}^{(S)} + \hat{\Gamma}^{(A)}$
be an additive conserved quantity for the composite system $S + A$, and $U$
the unitary evolution operator describing the system-apparatus interaction.
 As already remarked, the assumption of ideality of the measurement can be 
 expressed by the equation:

\begin{equation}
U[|m\rangle\otimes|A_{0}\rangle]\quad =\quad |m\rangle\otimes|A_{m}\rangle
\label{ghir30}
\end{equation}

while the fact that $\hat{\Gamma}$ is a conserved quantity implies:

\begin{equation}
U^{\dagger}\hat{\Gamma}U\quad =\quad\hat{\Gamma}. \label{ghir31}
\end{equation}

One can then argue along the following lines:

\begin{eqnarray}
& & \langle m'|[\hat{\Gamma}^{(S)}, \hat{M}]|m\rangle =
    \langle m'A_{0}|[\hat{\Gamma}^{(S)}, \hat{M}]|A_{0}m\rangle =
    \langle m'A_{0}|[\hat{\Gamma}, \hat{M}]|A_{0}m\rangle \nonumber \\
& & (m - m')\langle m'A_{0}|\hat{\Gamma}|A_{0}m\rangle =
    (m - m')\langle m'A_{0}|U^{\dagger}\hat{\Gamma}U|A_{0}m\rangle = \nonumber 
\\
& & (m - m')\langle m'A_{m'}|\hat{\Gamma}|A_{m}m\rangle = \label{ghir32} \\
& & (m - m')[\langle m'|m\rangle\langle A_{m'}|\hat{\Gamma}^{(A)}|A_{m}\rangle
    + \langle A_{m'}|A_{m}\rangle\langle m'|\hat{\Gamma}^{(S)}|m\rangle].
    \nonumber
\end{eqnarray}

	The last line shows that $\langle A_{m'}|A_{m}\rangle = 0$ 
for all $m\neq
m'$ implies $\langle m'|[\hat{\Gamma}^{(S)}, \hat{M}]|m\rangle = 0, \,\, \forall
m',m$, i.e., $[\hat{\Gamma}^{(S)}, \hat{M}] = 0$. Therefore, the 
observable $\hat{M}$ of $S$ must commute with the term  
$\hat{\Gamma}^{(S)}$ of the additive conserved 
quantity. It goes without saying that even in the very elementary case 
of the measurement of the spin component $\hat{S}_{z}$
of a spin $1/2$ particle, the 
ideal measurement scheme cannot hold because such a component does not 
commute with the spin component $\hat{S}_{x}$, which is a term of a conserved 
additive quantity (i.e. the component $\hat{J}_{x}$ of the total angular 
momentum). 
In ref. \cite{ghi15} the most general proof of this theorem has been derived 
in a completely rigorous way also for the case of unbounded operators.
Which lesson can we derive from the above analysis? Obviously, there 
are only two alternatives:

--- Either the measurement is distorting (i.e. the final state of the system is 
different from the initial one and final states corresponding to initial 
orthogonal states are no longer orthogonal --- at the r.h.s. of Eq. 
(\ref{ghir30})
$|m\rangle$ has to be replaced by another state $|\tilde{m}\rangle$
and it must hold $\langle\tilde{m}|\tilde{m}'\rangle \neq 0$) 
but then the apparatus 
cannot be used to ``prepare'' a system and, at any rate, one cannot make precise 
claims about the properties possessed by the system after the measurement, in 
particular they may differ from those which have been revealed by the apparatus, 

--- Or the final apparatus states cannot be orthogonal.

	Actually, as it has been discussed, e.g., in \cite{ghi13, ghi15}, 
realistic measurement processes are both distorting and lead to non-mutually 
exclusive outcomes. It is also well known \cite{ghi11, ghi12, ghi13, ghi14,
ghi15} that the amount of 
nonideality can be made arbitrarily small by making arbitrarily large, 
in the initial apparatus state, the mean value of the square of the 
conserved quantity. This nice feature is accepted by everybody as 
showing that the limitations induced by the presence of additive 
conserved quantities are not physically relevant and/or puzzling. This 
position contrasts with the author's absolutely strict request of 
ideality of the measurement.

\section{Conclusions.}

	The conclusions of our analysis should be obvious: one can have many 
reasons to dislike dynamical reduction models and to prefer other ways 
to overcome the difficulties of SQM like, e.g., adopting the point of 
view of Bohmian Mechanics (see however the remarks in ref. \cite{ghi16}), but 
certainly arguments like those put forward in ref.\cite{ghi1} have not to be 
taken into account in judging the appropriateness of the considered 
approach.

\section*{Acknowledgements.}

	We acknowledge useful suggestions by Prof. Detlef D\"{u}rr.


\begin{thebibliography}{99}

\bibitem{ghi1} P.J. Lewis, Brit. J. Phil. Sci., {\bf48}, 313 (1997).

\bibitem{ghi2} G.C. Ghirardi, A. Rimini and T. Weber, Phys. Rev. D {\bf34}, 
470 (1986).

\bibitem{ghi3} A. Shimony, in: {\it PSA 1990}, Vol.2, A. Fine, M. Forbes and 
L. Wessel eds., Philosophy of Science Association, East Lansing, Michingan, 1991
; D.Z. Albert and B. Loewer, in: {\it PSA 1990}, Vol.1, A. Fine, M. Forbes and 
L. Wessel eds., Philosophy of Science Association, East Lansing, Michingan, 
1991.

\bibitem{ghi4} G.C. Ghirardi, P. Pearle and A. Rimini, Phys. Rev. A {\bf42},
78 (1990).

\bibitem{ghi5} G.C. Ghirardi, R. Grassi and F. Benatti, Found. Phys.,
{\bf25}, 5 (1995).

\bibitem{ghi6} J.S. Bell, in {\it Schr\"{o}dinger-Centenary Celebration of a 
Polymath}, C.W. 
Kilmister ed., Cambridge University Press, Cambridge, 1987.

\bibitem{ghi7} P. Pearle, Phys. Rev., A {\bf39}, 2277 (1989); G.C. Ghirardi 
and A. Rimini, in {\it Sixty-Two Years of Uncertainty}, A. Milled ed. Plenum, 
New York, 1990; G.C. Ghirardi, R. Grassi and A. Rimini, Phys. Rev. A {\bf42}, 
1057 (1990); G.C. Ghirardi, R. Grassi and P. Pearle, Found. Phys., {\bf20}, 
1271 (1990).

\bibitem{ghi8} J. von Neumann, {\it Mathematische Grundlagen der 
Quantenmechanik}
(Springer, Berlin, 1932) (English edition: Princeton University Press,  
Princeton, 1955).

\bibitem{ghi9} E. Schroedinger, Naturwissenschaften, 23, 807-812; 823-828; 
844-849, (1935).

\bibitem{ghi10} A. Einstein, {\it Reply to criticisms}, in Schilpp, {\it Albert
Einstein Philosopher-Scientist}, The Open Court Publishing Co. La Salle, 
Illinois, 1949.

\bibitem{ghi11} E.P. Wigner, Z. Phys., 131, 101 (1952); H. Araki and M.M. 
Yanase, Phys. 
Rev. 120, 662 (1960); H. Stein and A. Shimony, in: {\it Foundations of Quantum 
Mechanics}, B. d'Espagnat ed., Academic Press, New York, 1971; M.M. Yanase in: 
{\it Foundations of Quantum Mechanics}, B. d'Espagnat ed., Academic Press, 
New York, 1971.

\bibitem{ghi12} G.C. Ghirardi, F. Miglietta, A. Rimini and T. Weber, Phys. 
Rev. D {\bf24}, 347 (1981).

\bibitem{ghi13} G.C. Ghirardi, F. Miglietta, A. Rimini and T. Weber, Phys. 
Rev. D {\bf24}, 353 (1981).

\bibitem{ghi14} G.C. Ghirardi, A. Rimini and T. Weber, J. Math. Phys., 
{\bf23}, 
1792 (1982).

\bibitem{ghi15} G.C. Ghirardi, A. Rimini and T. Weber, J. Math. Phys., 
{\bf24}, 2454 (1983).

\bibitem{ghi16} E. Deotto and G.C. Ghirardi, Found. Phys., {\bf28}, 1 (1998).

\end{thebibliography}
\end{document}